\documentclass[conference]{IEEEtran}
\usepackage{amsmath, graphicx, algorithm, algorithmic}

\begin{document}

\title{Deep Neural Network Based Resource Allocation for V2X Communications}

\author{\IEEEauthorblockN{Jin Gao$^1$,
        Muhammad R. A. Khandaker$^1$,
        Faisal Tariq$^2$,
        Kai-Kit Wong$^3$ and
        Risala T. Khan$^4$}
\IEEEauthorblockA{
        $^1$School of Engineering and Physical Sciences, Heriot-Watt University, Edinburgh, United Kingdom\\
        $^2$James Watt School of Engineering, University of Glasgow, United Kingdom\\
        $^3$Department of Electronic and Electrical Engineering, University College London, United Kingdom\\
        $^4$Institute of Information Technology, Jahangirnagar University, Dhaka, Bangladesh\\
Corresponding e-mail: $\rm m.khandaker@hw.ac.uk$}}

\maketitle

\begin{abstract}

This paper focuses on optimal transmit power allocation to maximize the overall system throughput in a vehicle-to-everything (V2X) communication system. We propose two methods for solving the power allocation problem namely the weighted minimum mean square error (WMMSE) algorithm and the deep learning-based method. In the WMMSE algorithm, we solve the problem using block coordinate descent (BCD) method. Then we adopt supervised learning technique for the deep neural network (DNN) based approach considering the power allocation from the WMMSE algorithm as the target output. We exploit an efficient implementation of the mini-batch gradient descent algorithm for training the DNN. Extensive simulation results demonstrate that the DNN algorithm can provide very good approximation of the iterative WMMSE algorithm reducing the computational overhead significantly.

\end{abstract}

\begin{IEEEkeywords}
Machine learning, deep learning, deep neural network, V2X, V2V, power control,  resource allocation.
\end{IEEEkeywords}

%
\IEEEpeerreviewmaketitle

\section{Introduction}

Vehicle-to-infrastructure (V2I) and vehicle-to-vehicle (V2V) communications, also known as V2X communications together, have become a hot research topic especially since the emergence of the fifth-generation (5G) communication systems \cite{MBoban_VTMag}. With advent of autonomous driverless vehicles, transmission units mounted on vehicles will need to exchange massive amount of information signals including speed, traffic condition, direction, location, traffic incidents etc. at the frequency of ten times or even more every second through high-speed wireless links. For example, when a car pulls an emergency breaking due to an unexpected emergency situation, the signal should be transmitted to surrounding vehicles to make them aware of potential hazard buildup in the vicinity. With the deployment of 5G wireless communication technology, it is no surprise that the vehicles can communicate with other communication devices, such as mobile phones and smart computers or even facilities like building, traffic lights and road and so on, which will enhance the proliferation and security of autonomous driving in the future \cite{6g_vision, what_will_5g_be}. 

The V2X communication systems feature several unique characteristics as opposed to conventional cellular communications including high mobility, rapid change of direction as well as location, and stringent quality-of-service (QoS) requirements. Furthermore, road safety concerns impose very strict requirements on ultra-low latency and high reliability in V2X communications. However, due to high mobility, the V2X communication channel state information (CSI) becomes outdated quite rapidly. Thus faster optimization techniques are of paramount interest for reliable V2X communications.

Furthermore, in order to improve the reliability of V2X communications, proper interference management as well as resource allocation strategies must be in place for the V2V and V2I links. Traditional interference and resource management schemes for V2X interference-limited scenarios operate in iterative manner, which has high computational complexity \cite{d2d_v2v, d2d_v2x}. While these conventional approaches provide a good understanding of the problem domain and reveals valuable insights of the V2X systems, these are unsuitable for most practical V2X systems requiring ultra-reliable low-latency communications (URLLC) \cite{what_will_5g_be, 6g_vision}.

Recently, machine learning approaches have gained momentum in the wireless communications domain due to their inherent capability of efficiently dealing with large-scale problems. Another reason for the boom of machine learning across a wide range of application domains is that machine learning approaches can combine learning process with existing field technologies. Through appropriate training process, the knowledge can generally decide on a particular hypothesis class. Wireless channel estimation and resource allocation problems are potential examples where machine learning are increasingly exploited \cite{jiang2016machine}. Of particular interest are deep learning techniques enabled by deep neural networks (DNN) due to their reduced computation time. Once trained properly, DNN can provide real-time resource allocation solutions, which is very crucial for V2X communications \cite{zhang2019deep}.

In \cite{drl_v2v_conf, drl_v2x_jrnl}, the authors have extended the works in \cite{d2d_v2x} by solving the same resource allocation problem using deep reinforcement learning (DRL) technique. In particular, the authors in \cite{drl_v2v_conf, drl_v2x_jrnl} developed a decentralized resource allocation technique for V2V communications based of DRL technique. However, with cloud radio access network (C-RAN) being an inherent part of 5G, it is generally expected that radio resource management strategies will be implemented at the centralized cloud \cite{what_will_5g_be}. The main advantage of C-RAN is that with access to enormous information in the cloud, radio resources can be allocated more efficiently. 

In this paper, we propose a centralized resource allocation strategy using deep learning technique for V2X communications, as opposed to the decentralized approaches in \cite{drl_v2v_conf, drl_v2x_jrnl}. In particular, we consider a sum rate maximization problem under individual power constraints for each V2I and V2V communication link using supervised learning technique. The main contributions in this paper are listed below:
\begin{itemize}
    \item a) We first develop an iterative power allocation algorithm for the proposed V2X communications system.
    \item b) We then propose a DNN based power allocation scheme for the V2V and the V2I links. We apply the mini-batch gradient descent (MBGD) algorithm by identifying the suitable batch size and learning rate through cross validation.
    \item c) We have validated the proposed learning algorithm through extensive simulations that demonstrate the suitability of the algorithm particularly for V2X communications.
\end{itemize}


\begin{figure}[ht!]
\includegraphics[width=\linewidth]{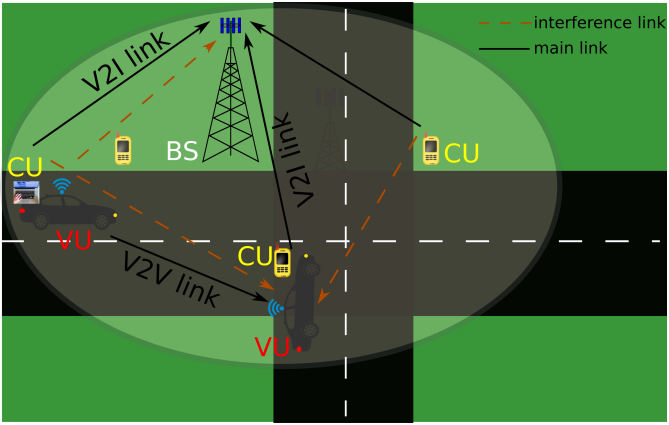}
\caption{The proposed V2V and V2I communication system.} \label{sys_mod}
\end{figure}

\section{System Model}\label{sec_vision}
Let us consider a high-mobility vehicular communication system (cf. Fig.~\ref{sys_mod}) in which $K$ pairs of vehicular nodes share high-capacity V2I communication links with $M$ cellular users (CUs) in a 6G communication system. For the ease of exposition, we consider single-antenna communication nodes. We also assume that all vehicles may accommodate both CUs and VUs simultaneously and are capable of radio transmissions from separate antennas. The CUs communicate with the roadside infrastructure which may be a cellular base station (BS), traffic lamppost, building or any other fixed structure. We also assume that the vehicles travel at a constant speed along the roads and data is transmitted at each time slot using the same carrier frequency.

The channel power gain between the $m$th CU and the BS is defined as \cite{d2d_v2x}
\begin{align}
    h_{m}^{\rm c} = G_{m}^{\rm c}\beta_{m}^{\rm c}Ad_{m}^{{\rm c}^{-\gamma}}, \label{ch_gain}
\end{align}
where $G_{m}^{\rm c}$ is assumed to be exponentially distributed fast fading power gain, $\beta_{m}^{\rm c}$ is the log normal shadow fading component, $A$ is the constant pathloss, $d_{m}^{\rm c}$ is the distance between the $m$th CU and the BS, and $\gamma$ is the pathloss exponent. Similarly, we can define the channel $h_{k}^{\rm v}$ between the $k$th V2V transmitter and the BS as well as the $k$th V2V link $g_{k}^{\rm v}$. Thus the signal received at the BS from the $m$th CU can be expressed as
\begin{multline}
  y_m^{\rm c} = \underbrace{h_{m}^{\rm c}s_m^{\rm c}}_{\text{desired signal}} + \underbrace{\sum_{n=1, n \ne m}^{M} h_{n}^{\rm c} s_n^{\rm c}}_{\text{interference from the CUs}} + \!\!\! \underbrace{\sum_{k=1}^{K} h_{k}^{\rm v}s_k^{\rm v}}_{\text{interference from the VUs}}\\ + \underbrace{n_m}_{\text{noise}}, \label{rx_sig}
\end{multline}
where $s_i^{\rm x} \sim \mathcal{CN}(0,P_i^{\rm x}),~ x \in \{\rm c, v\},$ is the signal transmitted by the $i$th transmitter and $n_m \sim \mathcal{CN}(0,\sigma^2)$ is the additive white Gaussian noise component at the BS.
Thus the received signal to noise plus interference ratio (SINR) at the BS for the $m$th CU can be expressed as
\begin{equation}
    \gamma_m^{\rm c} = \frac{P_m^{\rm c}|h_{m}^{\rm c}|^2}{\sum_{n=1,n\ne m}^M P_n^{\rm c}|h_{n}^{\rm c}|^2 + \sum_{k=1}^K P_k^{\rm v}|h_{k}^{\rm v}|^2 + \sigma^2}. \label{sinr_c}
\end{equation}
Similarly, the received SINR in the $k$th V2V link is given by
\begin{equation}
    \gamma_k^{\rm v} = \frac{P_k^{\rm v}|g_{k}^{\rm v}|^2}{\sum_{l=1,l\ne k}^K P_l^{\rm v}|g_{k}^{\rm v}|^2 + \sum_{m=1}^M P_m^{\rm c}|h_{m,k}|^2 + \sigma^2}, \label{sinr_v}
\end{equation}
where $h_{m,k}$ is the interfering channel from V2I user $m$ to V2V receiver $k$.

\subsection{Problem Formulation}

Our aim is to maximize the overall system throughput of the V2X links by optimally allocating transmit power among all the users. Accordingly, we formulate the following optimization problem:
\begin{subequations}\label{prob_msr0}
\begin{align}
\max_{\{P_m^{\rm c}\},\{P_k^{\rm v}\}} ~&~ \sum_{m=1}^M \alpha_m \log\left(1 + \gamma_m^{\rm c} \right) + \sum_{k=1}^K \alpha_k \log\left(1 + \gamma_k^{\rm v} \right) \label{prob_msr0_o}\\
{\rm s.t.} ~&~ 0 \le P_m^{\rm c} \le P_{\rm max}^{\rm c}, \forall m, \label{prob_msr0_c1}\\
 ~&~ 0 \le P_k^{\rm v} \le P_{\rm max}^{\rm v}, \forall k, \label{prob_msr0_c2}\\
 ~&~ \alpha_m, ~ \alpha_k \ge 0, \forall m, \forall k,\label{prob_msr0_c3}
\end{align}
\end{subequations}
where $P_{\rm max}^{\rm c}$ and $P_{\rm max}^{\rm v}$ are the maximum transmit power budget of the CUs and the VUs, respectively, $\{\alpha_m\}, ~ \{\alpha_k\}$ are the weights, which determine the priority of the corresponding V2I and V2V links. Note that the problem \eqref{prob_msr0} is non-convex and hence the exactly optimal solution is non-trivial mainly due to the $\log(\cdot)$ in the objective. In the following, we will first develop an acceptable solution to the problem following traditional alternating approaches, and then following machine learning techniques.

\section{The Weighted MMSE Algorithm}\label{sec_wmmse}
In this section, we propose an alternative approach of solving problem \eqref{prob_msr0}. Instead of solving problem \eqref{prob_msr0} directly, we solve the equivalent WMMSE minimization problem, inspired by \cite{wmmse}. The equivalence is guaranteed by the well-known MMSE-SINR equality as proved in \cite{palomar_majorization}. Note that here we downscale the WMMSE algorithm originally proposed for MIMO systems in \cite{wmmse} to the equivalent SISO channels for convenience.

Since neural networks work only with real numbers, the absolute value terms in \eqref{sinr_c} and \eqref{sinr_v} representing channel power gains tend to be beneficial in the neural network based design in Section~\ref{sec_dnn}. However, in the conventional WMMSE algorithm, real channel power gains are also convenient for mathematical operations. Inspired by \cite{wmmse}, we assume that $v_i^{\rm x},~ x \in \{\rm c, v\},$ is the amplifier gain used to transmit signal by the $i$th transmitter,  while $u_i^{\rm x}$ is the receiver amplifier gain used to estimate the desired signal. Accordingly, the estimated real symbol $\hat{s}_m^{\rm c}$ is given by \cite{kay}
\begin{multline}
    \hat{s}_m^{\rm c} = u_m^{\rm c} \left(|h_{m}^{\rm c}|v_m^{\rm c}s_m^{\rm c} + \sum_{n\ne m}^M|h_{n}^{\rm c}|v_n^{\rm c} s_n^{\rm c} \right.\\
    \left.+ \sum_{k = 1}^K|h_{k}^{\rm v}|v_k^{\rm v} s_k^{\rm v} + n_m \right), \quad m = 1, \cdots, M. \label{s_hat}
\end{multline}
Thus the MSE of estimating $s_m^{\rm c}$ is given by \cite{wmmse, jrnl_mur1}
\begin{multline}
    \varepsilon_m^{\rm c} =  {\rm E}\left(\hat{s}_m^{\rm c} - s_m^{\rm c}\right)^2\\
    = \left|u_m^{\rm c} h_{m}^{\rm c}v_m^{\rm c} - 1\right|^2 + \sum_{n\ne m}^M\left|u_m^{\rm c}h_{n}^{\rm c}v_n^{\rm c} \right|^2\\
    + \sum_{k = 1}^K\left|u_m^{\rm c}h_{k}^{\rm v}v_k^{\rm v} \right|^2 + \sigma^2 |u_m^{\rm c}|^2, \quad m = 1, \cdots, M, \label{mse1}
\end{multline}
where ${\rm E}(\cdot)$ indicates the statistical expectation operation. Similarly, the MSE of estimating $s_k^{\rm v}$ is given by
\begin{multline}
    \varepsilon_m^{\rm v} = \left|u_k^{\rm v} h_{k}^{\rm v}v_k^{\rm v} - 1\right|^2 + \sum_{l\ne k}^K\left|u_k^{\rm v}h_{l}^{\rm v}v_l^{\rm v} \right|^2\\
    + \sum_{m = 1}^M\left|u_k^{\rm v}h_{m}^{\rm c}v_m^{\rm c} \right|^2 + \sigma^2 |u_k^{\rm v}|^2, \quad k = 1, \cdots, K. \label{mse1_v}
\end{multline}
Thus following the MMSE-SINR equality $\varepsilon_m = \frac{1}{1 + \gamma_m^{\rm c}}$ derived in \cite{palomar_majorization}, the weighted sum rate (WSR) maximization problem \eqref{prob_msr0} can be equivalently expressed as the following WMMSE minimization problem
\begin{subequations}\label{prob_mmse1}
\begin{align}
\min_{\{u_i^{\rm x}\},\{v_i^{\rm x}\},\{w_i^{\rm x}\}} ~&~ \sum_{x \in \{\rm c, v\}}\sum_i \left(w_i^{\rm x}\varepsilon_i^{\rm x} - \log(w_i^{\rm x}) \right) \label{prob_mmse1_o}\\
{\rm s.t.} ~&~ 0 \le v_m^{\rm c} \le \sqrt{P_{\rm max}^{\rm c}}, \forall m, \label{prob_mmse1_c1}\\
 ~&~ 0 \le v_k^{\rm v} \le \sqrt{P_{\rm max}^{\rm v}}, \forall k, \label{prob_mmse1_c2}\\
 ~&~ \alpha_m, ~ \alpha_k \ge 0, \forall m, \forall k,\label{prob_mmse1_c3}
\end{align}
\end{subequations}
where $w_i^{\rm x}$ is a positive weight factor.

The WMMSE minimization problem \eqref{prob_mmse1} can be solved using block coordinate descent (BCD) method \cite{boyd}. In each phase, the BCD method optimizes one set of variables while fixing the rest. Checking the first optimality condition, one can find the optimal weight $w_i^{\rm x} = \frac{1}{\varepsilon_i^{\rm x}}$ in closed form and the optimal receiver gain as the well-known Wiener filter \cite{icspcs10}:
\begin{align}
    u_i^{\rm x} = \frac{h_{i}^{\rm x}v_i^{\rm x}}{\sum_{m=1}^M|h_{m}^{\rm c}|^2|v_m^{\rm c}|^2 + \sum_{k=1}^K|h_{k}^{\rm v}|^2|v_k^{\rm v}|^2 + \sigma^2}.\label{rx_gain}
\end{align}
In order to obtain the optimal transmitter gain $v_i^{\rm x}, ~ x \in \{\rm c, v\}$, we can decouple problem \eqref{prob_mmse1} for each transmitter, fixing the other variables, as
\begin{subequations}\label{prob_mmse2}
\begin{align}
\min_{\{v_i^{\rm x}\}} ~&~ \sum_i \left(w_i^{\rm x}\varepsilon_i^{\rm x} - \log(w_i^{\rm x}) \right) \label{prob_mmse2_o}\\
{\rm s.t.} ~&~ 0 \le v_i^{\rm x} \le \sqrt{P_{\rm max}^{\rm x}}. \label{prob_mmse2_c1}
\end{align}
\end{subequations}
Applying the Lagrangian multiplier approach \cite{boyd}, problem \eqref{prob_mmse2} can be solved for  $v_i^{\rm x}$. Considering the first-order optimality condition from the Lagrangian approach, we obtain the optimal $v_i^{\rm x}$ as
\begin{align}
    v_i^{\rm x} = \frac{\alpha_i^{\rm x}w_i^{\rm x}u_i^{\rm x}|h_i^{\rm x}|}{\sum_{x \in \{\rm c, v\}}\sum_i\alpha_i^{\rm x}w_i^{\rm x}|u_i^{\rm x}|^2|h_{i}^{\rm x}|^2 + \mu_i^{\rm x}\sigma^2}.\label{tx_gain}
\end{align}
Here $\mu_i^{\rm x} \ge 0$ is the Lagrange multiplier, which should chosen such that the complementary slackness condition on the power constraint \eqref{prob_mmse2_c1} is satisfied. The overall WMMSE procedure of solving problem \eqref{prob_mmse1} is summarized in Algorithm~\ref{alg_conv}. 
It has been proven in \cite{d2d_v2x} that the WMMSE algorithm eventually converges to a stationary point.

\begin{algorithm} [ht]
    \caption{WMMSE algorithm for solving problem \eqref{prob_mmse1}}\label{alg_conv}
  \begin{algorithmic}[1]
    \STATE Initialize $v_i^{\rm x}(0), ~ x \in \{\rm c, v\}$ such that $0 \le v_i^{\rm x} \le \sqrt{P_{\rm max}^{\rm x}}$.
    \STATE Compute $u_i^{\rm x}(0) = \frac{h_{i}^{\rm x}v_i^{\rm x}(0)}{\sum_{x \in \{\rm c, v\}}\sum_i|h_{i}^{\rm x}|^2|v_i^{\rm x}(0)|^2 + \sigma^2}$.
    \STATE Compute $w_i^{\rm x}(0) = \left[1 - u_i^{\rm x}(0)|h_{i}^{\rm x}||v_i^{\rm x}(0)|\right]^{-1}$.
    
    \STATE Set $n:=0$.
    
	\STATE \textbf{repeat}
      \STATE  $v_i^{\rm x}(n+1) \leftarrow \frac{\alpha_i^{\rm x}w_i^{\rm x}(n)u_i^{\rm x}(n)|h_i^{\rm x}|}{\sum_{x \in \{\rm c, v\}}\sum_i\alpha_i^{\rm x}w_i^{\rm x}(n)|u_i^{\rm x}(n)|^2|h_{i}^{\rm x}|^2}$.
	  \STATE  $u_i^{\rm x}(n+1) \leftarrow \frac{h_{i}^{\rm x}v_i^{\rm x}(n+1)}{\sum_{x \in \{\rm c, v\}}\sum_i|h_{i}^{\rm x}|^2|v_i^{\rm x}(n+1)|^2 + \sigma^2}$.
     \STATE $w_i^{\rm x}(n+1) \leftarrow \left[1 - u_i^{\rm x}(n+1)|h_{i}^{\rm x}||v_i^{\rm x}(n+1)|\right]^{-1}$.
     \STATE $n:= n + 1$.
    \STATE \textbf{until} convergence.
    \STATE \textbf{Output:} Optimal power profile $P_i^{\rm x} = v_i^{\rm x}(n))^2$.
  \end{algorithmic}
\end{algorithm}

\section{Proposed Machine Learning Approach}\label{sec_dnn}
It has been shown in \cite{drl_v2v_conf, drl_v2x_jrnl} that a multi-layer neural network (MLNN) can provide a very good approximation of the WMMSE algorithm to leverage the computational efficiency of the DNN. Hence in the following, we propose a deep learning based power allocation scheme for the WSR maximization problem \eqref{prob_msr0}. The proposed DNN algorithm offers multi-fold benefits compared to the WMMSE approach. 

The DNN algorithm operates by continuous mapping of values from the iterations of the WMMSE algorithm.  In other words, with the target output from the WMMSE algorithm, the DNN needs to learn and approximate the unknown relationship between the input and the output. Consequently, this `black box' is transformed to a nonlinear mapping that can perform like the WMMSE approach.

\begin{figure}[ht!]
\centering
\includegraphics[width=0.8\linewidth]{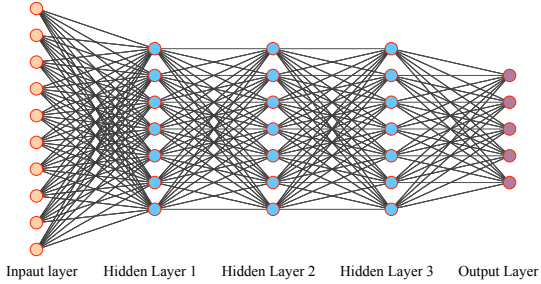}
\caption{The proposed deep neural network for approximating the WMMSE power control problem. Input: channel power gains, Output: optimal power.}\label{fig_dnn}
\end{figure}

\subsection{Defining the Neural Network}
 For the machine learning based power allocation scheme, we consider a supervised learning approach which approximates the iterative WMMSE algorithm using a fully connected neural network. The DNN consists of one input layer, multiple hidden layers and one output layer as shown in Fig.~\ref{fig_dnn}. While increasing the number of hidden layers may help in decreasing the number of hidden neurons at each hidden layer thus improving the computation efficiency, the optimal number of hidden layers is intractable. In our simulations (in Section~\ref{sec_sim}) we consider three hidden layers resulting in good performance efficiency. The channel power gains of the V2I and the V2V links are the inputs to the DNN and the optimal power allocation is the output of the DNN. Note that we do not consider any bias inputs to the neurons at any layer. We apply rectified liner unit (ReLU) as the activation function at the hidden layers, while the activation function at the output layer is specifically tailored to enforce the power constraints in problem \eqref{prob_msr0}. In particular, the hidden layer activation function is defined as
 \begin{align}
     y_{\text{hidden}} = \max(0,x_{\text{hidden}}), \label{act_hidden}
 \end{align}
 and that at the output layer is defined as
 \begin{align}
     y_{\text{out}} = \min\left(\max(0,x_{\text{out}}),P_{\max}^{\rm x}\right). \label{act_out}
 \end{align}

\subsection{Training the DNN}
For training the DNN, we first generate a large set of channel realizations $\left\{h_{m,{\rm B}}\right\}$ and $\left\{g_{k,{\rm B}}\right\}$ following certain channel distributions to reflect the V2I and V2V channels as defined in \eqref{ch_gain}. Then we generate the corresponding power allocation for each training sample using the DNN in Fig.~\ref{fig_dnn}. The training data sets are used to optimize the weights of the neural network such that the MSE of the target power allocation obtained from the WMMSE algorithm and the current DNN output is minimized. Accordingly, we define the cost function as the MSE between DNN output and the target output as
\begin{align}
    MSE = \frac{1}{M + K}\sum_{i = 1}^{M + K}(\text{out}_i - \text{target}_i)^2.
\end{align}

Since training the DNN with a very large data samples is highly time-demanding, we use the so called mini-batch gradient descent (MBGD) algorithm for training the proposed DNN. An efficient implementation of the MBGD algorithm which divides the learning rate (or, gradient) by a running average of magnitudes of recent gradients. This approach is referred to as RMSprop \cite{rmsprop}. In RMSprop, the running average of the squared gradients is updated as
\begin{multline}
    E\left\{\left(\nabla_w(t)\right)^2\right\} = 0.9*E\left\{\left(\nabla_w(t-1)\right)^2\right\}\\ + 0.1*\left(\nabla_w(t)\right)^2, \label{rms_prop}
\end{multline}
and the weights ($w$) at iteration $t$ is updated as \cite{adaptive_sg}
\begin{multline}
    w(t+1) = w(t) - \frac{\eta}{\sqrt{E\left\{\left(\nabla_w(t)\right)^2\right\} + \epsilon}}\nabla_w(t), \label{rms_prop}
\end{multline}
where $E$ indicates arithmetic average, $\nabla_w(t) = \frac{\partial MSE}{\partial w(t)}$ is the gradient of the learning objective (in our case, the MSE) and $\epsilon$ is a smoothing term to avoid division by zero (usually on the order of $1e-8$) \cite{grad_desc}. Note that the running average at time $t$ in \eqref{rms_prop} depends only on the most recent average of the squared gradients (at time $t-1$) and the current gradient. Interestingly, the previous average squared-gradient carries more weight ($90\%$) than the current gradient ($10\%$). It has been shown in \cite{adaptive_sg} that dividing the gradient by the square root of the running average makes the learning work much better. In order to normalize the variance of each neuron's output, we then divide the weights of each neuron by the square root of is number of inputs. Similar tactic has also been applied in \cite{lto_conf}.

\subsection{Testing the DNN}
Once we learn the optimal weights of the neural network from the training stage, the next task is to validate the performance of the neural network based approach with testing data set. Again we generate a reasonably large number of random channel realizations following the same distribution as the did for training. Each set of channels is then applied to the input of the \textit{trained} neural network for an optimal power allocation at the output. The sum rate is then averaged over the number of test data sets.

\section{Numerical Simulations}\label{sec_sim}
In this section, we perform numerical simulations to demonstrate the effectiveness of the proposed machine learning based power allocation scheme for wireless V2X communications. Throughout this section, we compare the performance of the proposed approach against the WMMSE based iterative benchmark scheme, originally proposed for a MIMO system in \cite{wmmse}. Towards this end, we first simulate the equivalent WMMSE scheme developed in Section~\ref{sec_wmmse}, following Algorithm~\ref{alg_conv}.

The V2X network consists of one cellular base station, $M = 8$ CUs and $K = 10$ V2V transmitters. We construct a fully connected DNN for the system with one input layer, three hidden layers and one output layer. The input layer consists of $N+ K + N \times K$ neurons, the three hidden layers consist of $50, 22, 20$ neurons, respectively, and the output layer has $N+K$ neurons to produce the power profile. The V2X channels are generated following the model in \cite{d2d_v2x} and assumed to be exponentially distributed. In all cases, the noise variance is assumed to be $0.1$.

It is important to choose a suitable batch-size as well as the learning rate for the MBGD algorithm since both parameters can affect the network performance and efficiency. Unfortunately, there is no generic mathematical model for defining suitable values for these two parameters. Therefore, we choose these parameters by cross-validation during the training phase. 

\begin{figure}
\centering
\includegraphics[width=0.8\linewidth]{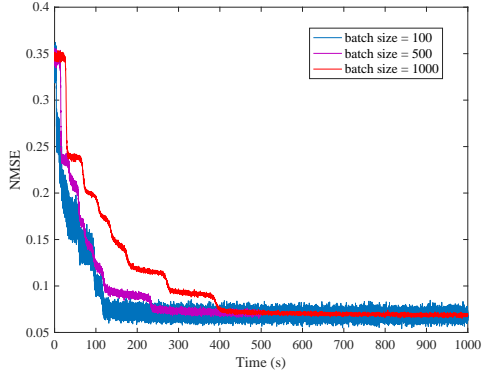}
\caption{Effect of batch size on learning time.}\label{bat_size}
\end{figure}

\begin{figure}
\centering
\includegraphics[width=0.8\linewidth]{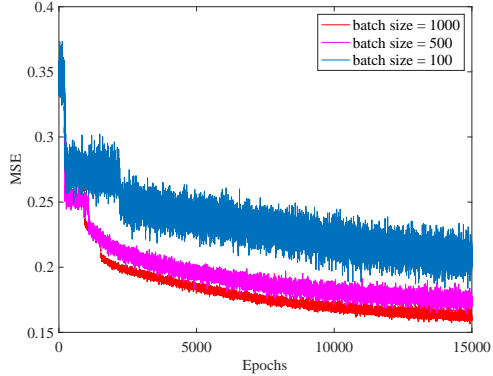}
\caption{Effect of batch size on training epoch.}\label{bat_epoch}
\end{figure}

The effects of `batch size' on learning time and the required number of epochs have been demonstrated in Figs.~\ref{bat_size} and \ref{bat_epoch}, respectively. In both pictures, the fluctuation of batch size = $100$ is much more significant than other two curves, while batch size 1000 experiences the smallest fluctuation. The results in Fig.~\ref{bat_size} show the training speed with batch size $100$ requiring the least amount of time to reach the convergence, while batch size $1000$ costs the most time. On the other hand, Fig.~\ref{bat_epoch} demonstrates that the batch size $1000$ requires the least number of epochs to converge, which is in contradiction with the results in Fig.~\ref{bat_size}. We must make our choice trading off the opposing tendencies of the two graphs. While the different batch size results in Fig.~\ref{bat_size} meet after a while, those in Fig.~\ref{bat_epoch} do not. Although the graphs in Fig.~\ref{bat_epoch} are different from each other, we must stop at some point, just like the results in Fig.~\ref{bat_size}. If the epoch extends to a very large value, the difference at the beginning may become hard to be distinguished, especially for batch size $500$ and batch size $1000$.

\begin{figure}
\centering
\includegraphics[width=0.8\linewidth]{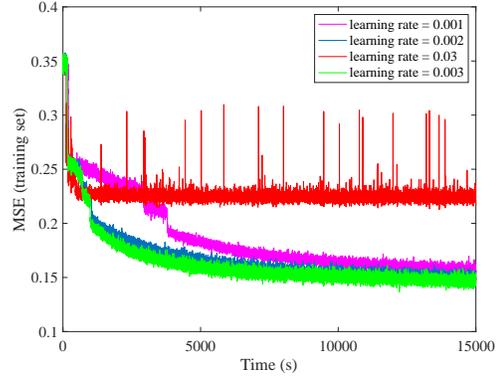}
\caption{Impact of learning rate on training MSE.}\label{lrate_all}
\end{figure}

The effect of learning rate on the training MSE has been shown in Fig.~\ref{lrate_all}. The results demonstrate that the cost function converges faster for larger learning rates at the low end, however, if the learning rate is significantly high (e.g., $0.03$), the MSE performance may not show a similar tendency and start worsening instead. Although intuitively a larger learning rate means faster converging speed, the learning rate cannot be increased arbitrarily as indicated by the results in Fig.~\ref{lrate_all}.


\begin{figure}
\centering
\includegraphics[width=0.8\linewidth]{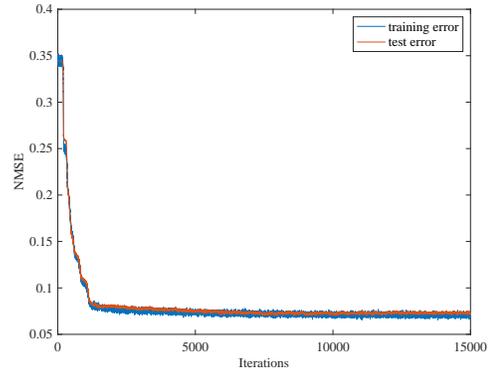}
\caption{Performance on the test data set. The data set is exclusively generated after the training phase.}\label{test_error}
\end{figure}

After parameter selection by cross-validation and improvement of DNN, we evaluate the performance of this trained network. In Fig.~\ref{test_error}, we compare the MSE performance of the test data set not exclusively used in training the DNN against the training data set. The blue curve is the MSE during training the DNN while the red one denotes to the testing error. From Fig.~\ref{test_error}, it is obvious that:
a) no `over fitting' problem occurs, and the gap between training error and testing error is negligible,
b) the direction of convergence on the target data set is as desirable since there is no big fluctuation,
c. the performance of the DNN is determined by its testing error and the performance in this case is acceptable since the error is very low.

\begin{figure}
\centering
\includegraphics[width=0.8\linewidth]{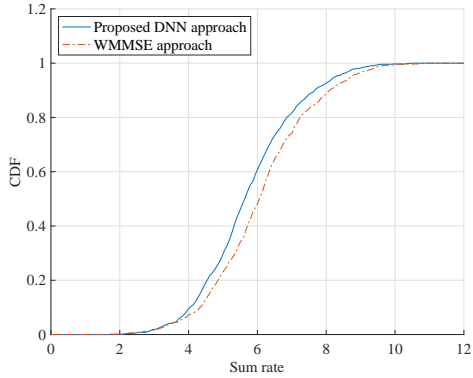}
\caption{The CDF of the achievable sum rates for the proposed DNN approach and the WMMSE-based baseline scheme.}\label{v2x_cdf}
\end{figure}

\begin{figure}
\centering
\includegraphics[width=0.8\linewidth]{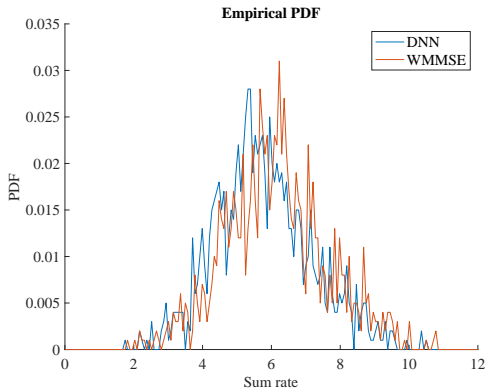}
\caption{The probability density function (PDF) of the proposed scheme against the WMMSE scheme.}\label{v2x_pdf}
\end{figure}

Next, we evaluate the sum rate performance of the proposed DNN-based power allocation scheme against the WMMSE scheme proposed in \cite{wmmse} originally for a conventional interference system. Fig.~\ref{v2x_cdf} shows the cumulative distribution function (CDF) of the two schemes for $10^5$ random test samples. Since we adopted the supervised learning technique, the DNN approach can not outperform the WMMSE algorithm. However, it can be easily observed from Fig.~\ref{v2x_cdf} that the proposed scheme has comparable sum rate performance against the WMMSE approach. 

Finally, we plot the probability density functions (PDFs) of the sum rate for both algorithms in Fig.~\ref{v2x_pdf}. The result of the trained network and the WMMSE algorithm is shown by their PDFs. From Fig.~\ref{v2x_pdf}, it is obvious that the DNN approach follows almost identical distribution to the WMMSE approach. These results further validates the effectiveness of the proposed machine learning approach.

\section{Conclusion}\label{sec_con}
We have considered the resource allocation problem for wireless V2X communications and shown that deep learning techniques bring enormous prospects for V2X communications where we need fast processing of optimization algorithms. Although training the DNN is a time-consuming procedure, the trained network can, however, offer a much faster solution. We have applied supervised learning approach to develop a DNN based power allocation scheme taking the classical iterative WMMSE based solution as the base line. While the  supervised learning can not offer a better solution than the target one, our results demonstrate that the proposed DNN based solution can achieve very close performance to the WMMSE approach. A particular aspect of the proposed solution is that the trained DNN offers a real-time solution which is crucial for V2X communications. However, developing an unsupervised learning based power allocation scheme could be an interesting future work. 

\label{sec4}

\ifCLASSOPTIONcaptionsoff
  \newpage
\fi

\bibliographystyle{IEEEtran}\footnotesize{

\bibliography{IEEEabrv,refdb}}%

\begin{thebibliography}{10}
\providecommand{\url}[1]{#1}
\csname url@samestyle\endcsname
\providecommand{\newblock}{\relax}
\providecommand{\bibinfo}[2]{#2}
\providecommand{\BIBentrySTDinterwordspacing}{\spaceskip=0pt\relax}
\providecommand{\BIBentryALTinterwordstretchfactor}{4}
\providecommand{\BIBentryALTinterwordspacing}{\spaceskip=\fontdimen2\font plus
\BIBentryALTinterwordstretchfactor\fontdimen3\font minus
  \fontdimen4\font\relax}
\providecommand{\BIBforeignlanguage}[2]{{%
\expandafter\ifx\csname l@#1\endcsname\relax
\typeout{** WARNING: IEEEtran.bst: No hyphenation pattern has been}%
\typeout{** loaded for the language `#1'. Using the pattern for}%
\typeout{** the default language instead.}%
\else
\language=\csname l@#1\endcsname
\fi
#2}}
\providecommand{\BIBdecl}{\relax}
\BIBdecl

\bibitem{MBoban_VTMag}
M.~Boban, A.~Kousaridas, K.~Manolakis, J.~Eichinger, and W.~Xu, ``Connected
  roads of the future: {Use} cases, requirements, and designconsiderations for
  vehicle-to-everything communications,'' \emph{IEEE Veh.Tech. Magazine},
  vol.~13, pp. 110--123, Sep. 2018.

\bibitem{6g_vision}
F.~Tariq, M.~R.~A. Khandaker, K.-K. Wong, M.~Imran, M.~Bennis, and M.~rouane
  Debbah, ``A speculative study on {6G},'' \emph{IEEE Commun. Magazine}, June
  2019 (submitted). Available: https://arxiv.org/pdf/1902.06700.pdf.

\bibitem{what_will_5g_be}
J.~G.~A. et~al., ``What will {5G} be?'' \emph{IEEE J. Sel. Areas Commun.},
  vol.~32, pp. 1065--1082, June 2014.

\bibitem{d2d_v2v}
W.~Sun, E.~G. Str\"{o}m, F.~Br\"{a}nnstr\"{o}m, K.~C. Sou, and Y.~Sui, ``Radio
  resource management for {D2D}-based {V2V} communication,'' \emph{IEEE Trans.
  Veh. Commun.}, vol.~65, pp. 6636--6650, Aug. 2016.

\bibitem{d2d_v2x}
L.~Liang, G.~Y. Li, and W.~Xu, ``Resource allocation for {D2D}-enabled
  vehicular communications,'' \emph{IEEE Trans. Commun.}, vol.~65, pp.
  3186--3197, July 2017.

\bibitem{jiang2016machine}
C.~Jiang, H.~Zhang, Y.~Ren, Z.~Han, K.-C. Chen, , and L.~Hanzo, ``Machine
  learning paradigms for next-generation wireless networks,'' \emph{IEEE
  Wireless Commun.}, vol.~24, pp. 98--105, 2016.

\bibitem{zhang2019deep}
C.~Zhang, P.~Patras, and H.~Haddadi, ``Deep learning in mobile andwireless
  networking: {A} survey,'' \emph{IEEE Commun. Surveys \& Tutorials}, 2019.

\bibitem{drl_v2v_conf}
H.~Ye and G.~Y. Li, ``Deep reinforcement learning for resource allocation in
  {V2V} communications,'' in \emph{Proc. IEEE Int. Conf. Commun. (ICC)}, Kansas
  City, MO, 2018.

\bibitem{drl_v2x_jrnl}
H.~Ye, G.~Y. Li, and B.~F. Juang, ``Deep reinforcement learning based resource
  allocation for {V2V} communications,'' \emph{IEEE Trans. Veh. Technology},
  vol.~68, no. 3163-3173, Apr. 2019.

\bibitem{wmmse}
Q.~Shi, M.~Razaviyayn, Z.~Luo, and C.~He, ``An iteratively weighted {MMSE}
  approach to distributed sum-utility maximization for a {MIMO} interfering
  broadcast channel,'' \emph{IEEE Trans. Signal Process.}, vol.~59, pp.
  4331--4340, Sep. 2011.

\bibitem{palomar_majorization}
D.~P. Palomar and Y.~Jiang, \emph{{MIMO} Transceiver Design via Majorization
  Theory}.\hskip 1em plus 0.5em minus 0.4em\relax now Publishers, 2007.

\bibitem{kay}
S.~M. Kay, \emph{Fundamentals of Statistical Signal Processing: Estimation
  Theory}.\hskip 1em plus 0.5em minus 0.4em\relax Englewood Cilffs, NJ:
  Prentice Hall, 1993.

\bibitem{jrnl_mur1}
M.~R.~A. Khandaker and Y.~Rong, ``Joint transceiver optimization for multiuser
  {MIMO} relay communication systems,'' \emph{IEEE Trans. Signal Process.},
  vol.~60, pp. 5977--5986, Nov. 2012.

\bibitem{boyd}
S.~Boyd and L.~Vandenberghe, \emph{Convex Optimization}.\hskip 1em plus 0.5em
  minus 0.4em\relax Cambridge, U.~K.: Cambridge University Press, 2004.

\bibitem{icspcs10}
M.~R.~A. Khandaker and Y.~Rong, ``Joint source and relay optimization for
  multiuser {MIMO} relay communication systems,'' in \emph{Proc. 4th Int. Conf.
  Signal Process. Commun. Systems (ICSPCS'2010)}, Gold Coast, Australia, Dec.
  13-15, 2010.

\bibitem{rmsprop}
G.~Hinton, N.~Srivastava, and K.~Swersky, ``Lecture 6a overview of mini-batch
  gradient descent,'' {Coursera} Lecture Slides, 2012. [Online]. Available:
  https://class.coursera.org/neuralnets-2012-001/lecture.

\bibitem{adaptive_sg}
J.~Duchi, E.~Hazan, and Y.~Singer, ``Adaptive subgradient methods for online
  learningand stochastic optimization,'' \emph{J. Machine Learning Research},
  vol.~12, pp. 2121--2159, Jan. 2011.

\bibitem{grad_desc}
S.~Ruder, ``An overview of gradient descent optimization algorithms,''
  \emph{arXiv preprint. Available: arXiv:1609.04747}, June 2017.

\bibitem{lto_conf}
H.~Sun, X.~Chen, Q.~Shi, M.~Hong, X.~Fu, and N.~D. Sidiropoulos, ``Learning to
  optimize: {Training} deep neural networks for wireless resource management,''
  in \emph{Proc. IEEE 18th Int. Workshop on Signal Process. Adv. Wireless
  Commun. (SPAWC)}, Sapporo, 2017.

\end{thebibliography}

\end{document}